\documentclass[twocolumn,showpacs,preprintnumbers,amsmath,amssymb]{revtex4}
\usepackage{dcolumn}% Align table columns on decimal point
\usepackage{graphics}
\usepackage{bm}% bold math
\begin{document}
%\preprint{APS/123-QED}
\title{Optical sorting and detection of sub-micron objects in a motional standing wave}
\author{Tom\'{a}\v{s} \v{C}i\v{z}m\'{a}r,  Martin \v{S}iler, Mojm\'{\i}r \v{S}er\'{y}, Pavel Zem\'{a}nek}  \email{zemanek@isibrno.cz}
\affiliation{Institute of Scientific  Instruments,
Academy of Sciences of the Czech Republic,\\
Kr\'{a}lovopolsk\'{a} 147, 612 64 Brno, Czech Republic}

\author{Veneranda Garc\'{e}s-Ch\'{a}vez, Kishan Dholakia}
\affiliation{School of Physics and Astronomy, University of St.  
Andrews, North Haugh, Fife, KY16 9SS, Scotland}

\date{\today}
\begin{abstract}
An extended interference pattern close to surface may result in both a
transmissive or evanescent surface fields for large area manipulation of trapped
particles. 
The affinity of differing particle sizes to a moving standing wave light
pattern allows us to hold and deliver them in a bi-directional
manner and importantly demonstrate
experimentally particle sorting in the sub-micron region.
This is performed without the need of fluid flow (static sorting).
Theoretical calculations experimentally confirm that
certain sizes of colloidal
particles thermally hop more easily between neighboring traps. A new generic
method is also presented for
particle position detection in an extended periodic light pattern and applied to characterization of optical traps and particle behavior. 

\end{abstract}

% PACS, the Physics and Astronomy
                             % Classification Scheme.

\maketitle
 
Optical trapping and guiding of micron and sub-micron size
objects have been a key topic of numerous studies in
the last three decades in the realm of mesoscale science
\cite{LangAJP03}. 
A recent drive in the field is the ability to interact with large scale
ensembles of particles in two-dimensions (2D) and ultimately in three-dimensions (3D): interferometric patterns may
help achieve this.
Light fields at or near total internal reflection may
assist allowing one to organize upwards of 1000 particles adjacent to a  surface
\cite{GuAPL04,Garces-ChavezAPL05}.
Recent work has shown the experimental
demonstration of optical separation using an extended
optical lattice
or holographic methods  in the presence
of a laminar flow \cite{MacDonaldNAT03,KordaPRL02} and it has been followed by theoretical analyses \cite{LacastaPRL05}. 
In these studies the particles are not
trapped {\em per se} but rather their differing affinity to a periodic
light pattern were exploited.
This sensitivity is very high and potentially offers a new non-invasive method for optical separation or sorting. 
Thus it would be a 
key advance to show how 
we may separate sub-micrometer objects in the absence of a flow 
and potentially over a large area, facilitating greater throughput.
The near field optical trapping has shown the ability to potentially organise particles over areas of $\mathrm{mm^2}$ and offers a potential  test area for sorting over a large region 
\cite{GuAPL04,Garces-ChavezAPL05,QuidantOL05}.
Tuning the incident angle from below 
to above the critical angle results in a transition between propagating 
and evanescent light fields. 
This geometry for such optical organization\cite{Garces-ChavezSPIE04} is eminently suitable for sorting, even if 
a small part of the incident focused beam is below the critical angle and thus forms propagating field. 
Ultimately we are creating light field patterning in 2D over an extended region.

In this letter we present an interferometric system that 
provides confinement and two-directional controllable delivery 
of particles of sub-micrometer sizes. 
Our system is employed in a geometry that both permits evanescent wave confinement as well as confinement just above the surface (by judicious choice of incident angle): this choice of geometry in all instances allows extension of the data presented here to a large area and concomitant higher throughput. 
The key result we show is the optical separation of sub-micrometer particles in the absence of any imposed flow by exploiting the varying affinity of objects to this spatially modulated light pattern. 
We are able to separate objects based upon their size. 
Additionally we demonstrate an original method for particle detection within this periodic near field light pattern
that is applied here for quantitative study of sub-micron particle behavior but can serve generically in any area where scattering of imposed periodic light pattern by objects on a surface can be detected (interaction of colloids, colloidal dynamics, colloidal models of thermodynamic systems, optical manipulation and sorting). 

The interference of two counter-propagating (C-P) surface fields (combination of evanescent and propagating waves) at or near the critical angle creates standing waves 
(SW) above or at the surface. As our data applies for both these cases we generally term this effect a surface standing wave field 
(SSW).  
In contrast to the previous methods \cite{GuAPL99} 
we used a setup with two independent C-P 
Gaussian beams that were focused on the top surface of the prism. 
In comparison to the single beam case of the same intensity, the interferometric optical potential well is deeper and steeper and consequently gives stronger longitudinal force component \cite{ZemanekOPTCOMM03}.
 The end result is that smaller particles can be confined and transported rapidly in both longitudinal directions 
if the SSW is moved.
This motion can be induced by altering the phase of one of the incident waves (in our case by movable mirror), resulting in a surface 
optical conveyor belt (OCB).  
This is related to recent work in sliding SWs
using radiative fields that have been exploited for atom 
delivery using Gaussian beams 
\cite{ZemanekPhD,DotsenkoPRL05}
 or quite recently  for sub-micron particle delivery using Bessel beams \cite{CizmarAPL05}. 

\begin{figure}[th]                                                     
\scalebox{0.26}{\includegraphics{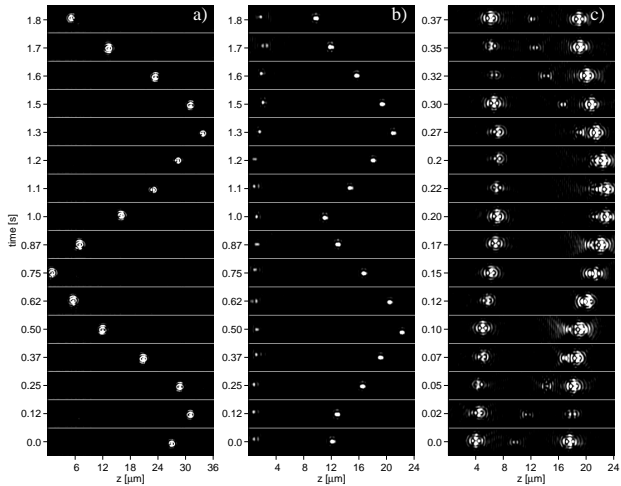}}
\caption{Proof of particle delivery in OCB but also particle affinity to OCB. a) A polystyrene sphere of diameter 520 nm is delivered over a distance of 36 $\mu$m. b) Mixture of polystyrene spheres of sizes 410 nm (left) and 520 nm (right). While the bigger moves with OCB, the smaller keeps its original position unaffected by OCB. c) Mixture of polystyrene particles of diameters 350 nm and two 750 nm where the smaller one follows the OCB motion but bigger spheres stay unaffected till the moment the smaller particle approached it. But one can notice that even before their physical contact there is a noticeable movement of the  bigger sphere caused by the interactions between both objects. }
\label{OCB}
\end{figure}

In this manner we were able to move a range of particles of sizes 350, 520, 600, 700 and 800~nm
over  a range of 40~$\mu$m on the surface (Fig \ref{OCB}a).   
However we found that beads of size 410~nm and 750~nm in contrast did NOT follow the moving SSW 
and remained unaffected by the imposed SSW 
(see Fig. \ref{OCB}b, \ref{OCB}c). 
Tuning the incident angle may result in either a surface patterned evanescent
fields or a propagating light field just above the surface. 
%Fig. 1

\begin{figure}[bpt]                                                               
\scalebox{0.45}{\includegraphics{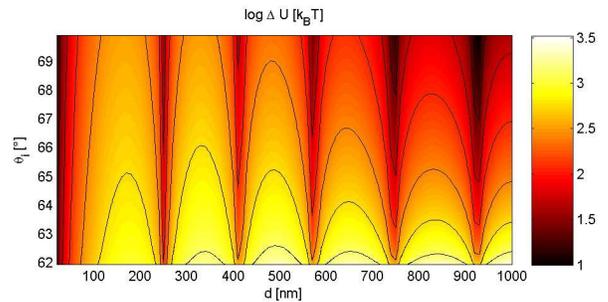}}
\caption{Theoretical results showing how the longitudinal trap depth in the SW
formed by C-P 
near-fields depends on the diameter of the polystyrene sphere and on the incident angle of the 
plane waves (the same for both C-P 
waves) upon the prism surface. 
The trap depth is in logarithmic scale (see the bar for units). 
The contour 
lines show the trap depths of  33 $kT$, 100 $kT$, 333 $kT$, 1000 
$kT$ and 3333 $kT$.
Polarizations of both beams are normal to the plane of incidence, i.e. parallel 
with prism surface, 
the amplitude of a single plane wave electric field is the same as it is in the centre of 1W 
Gaussian beam 
focused to a beam waist $1 \mu$m ( $E = $ 200  $V m^{-1}$).
}
\label{dU}
\end{figure}

We examine this behavior by a theoretical study based on an optical force calculation using Lorenz-Mie scattering model 
\cite{AlmaasJOSAB95} but modified to the 
C-P 
evanescent waves. 
It does not include 
surface interactions for the scattered light \cite{LesterOL99} but it is to be noted that these mainly affect the force component perpendicular to the surface which is not the dominant consideration for the work presented here.  
Figure \ref{dU} shows the theoretical depths of the optical traps and illustrates the selectivity of the 
SSW
to various sizes of objects. 
The lighter regions indicate deep traps where the objects can be confined with smaller requirements on beam intensities. 
The dark region corresponds to a shallow trap and in accordance with the lower beam intensity the object stays here only shortly and due to thermal activation hops to a neighboring trap site. 
An important conclusion coming from this model is that the sphere sizes affected and unaffected by this phenomena do not strongly depend on the considered incident angle of the plane waves.
Experimentally our observations concur with the above theory for monodisperse (in size) or diluted samples
on the top of the prism and we can readily observe the differing affinity of the particles to the SW
pattern similar to Figs. \ref{OCB}b, \ref{OCB}c. 
But dense colloidal samples suffered from inter-particle interactions. 
An example of this can be seen in Fig. \ref{OCB}c when the insensitive bigger particle 
started to move even before the smaller one physically touched it. 
This is attributed to an optical binding type interaction \cite{BurnsPRL89,TatarkovaPRL02} where scattered light from the particle selected by the OCB 
affects the optical forces experienced by the unaffected particle size.

\begin{figure}[tp]                                                              
\scalebox{0.41}{\includegraphics{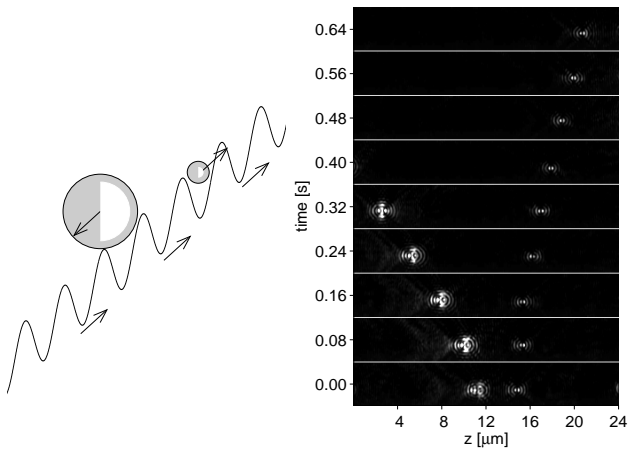}}
\caption{An example of a new method of optical sorting of colloids. The smaller sphere is 
delivered against the tilted potential landscape via OCB. The bigger particle is insensitive to the %SSW and therefore the radiation pressure coming from the right (in the direction of tilted potential) %accelerates the particle down the hill. The diluted sample was used here made of mixture of 
polystyrene particles of diameter 750 nm (left) and 350 nm (right). }
\label{sorter}
\end{figure}

Therefore to exploit the above described phenomena for separation of particles according to their size we exposed diluted samples
 to a tilted washboard potential \cite{TatarkovaPRL03}
 - this was created by introducing a slight intensity asymmetry in the two C-P 
beams. 
This asymmetry created a differential optical gradient along the SW %standing wave 
and if this was moved, 
it resulted in the particle behavior demonstrated in Fig. \ref{sorter}.
The same mechanism can be used also for 3D
 trapping over longer distance if the OCB 
based on non-diffracting beams is used \cite{CizmarAPL05}.

%Fig. 2
%Fig3
%Fig4
\begin{figure}[tp]                                                               
\scalebox{0.47}{\includegraphics{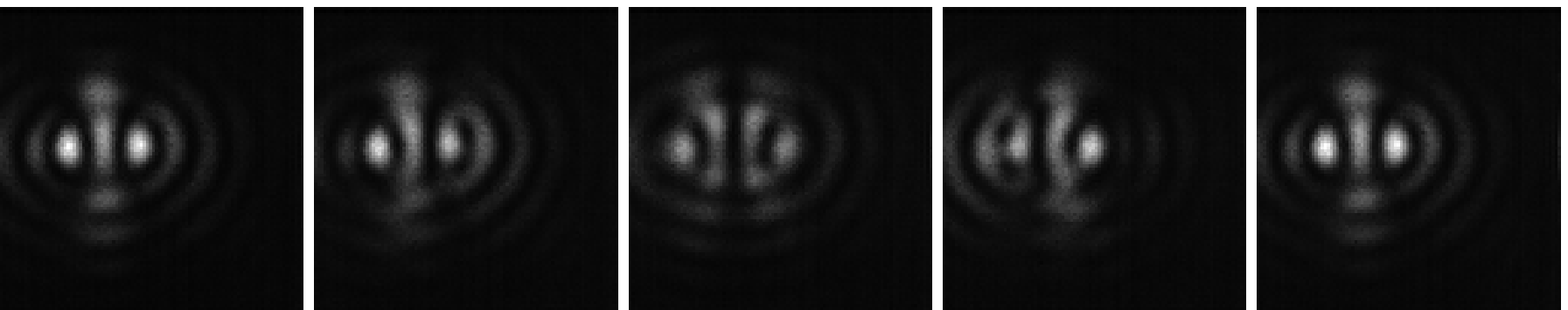}}
\caption{An example of five images of a 520 nm particle taken during fast sweep of the SSW over %the particle. 
The particle was assumed to be stationary and uninfluenced by the SSW motion. 
The first and the last images correspond to the particle equilibrium position. 
 }
\label{gallery}
\end{figure}

To explore in detail 
the properties of SSW
we developed an
original method of particle detection within this optical landscape.
The basis for this method lies in changes of the observed
light pattern created by the scattered light in the plane of CCD when the mutual 
configuration of the particle and SSW 
is changed  
(see Fig. \ref{gallery}). 
Since the 
SSW has a sinusoidal intensity dependence along the propagation
axis, we found parameters for a sinusoidal
dependence of intensity at each pixel of CCD
camera for different 
configuration of SSW and a particle. 
These functions enabled reconstruction of the interference pattern  
for each longitudinal position of a particle in the SSW.
Consequently as the particle moves in the fixed trap, the interference pattern is copying this motion around the CCD but in the axis of beam propagation this motion is linked with changes of the interference pattern shape as described above. 
If we want to determine the particle coordinates from the tracking record, we have to find for each frame the reconstructed pattern of the right shape placed to the right position using translation property of Fast Fourier Transform. 
This procedure gives us two information of longitudinal particle position - with respect to the CCD (from the image shift) and to the SSW (from the pattern shape). 
The later one can be obtained with nanometer resolution regardless of the stability of SSW with respect to the CCD. 
Unfortunately this is not generally true for the lateral direction which can only be found from the image shift. 
During
the particle tracking experiments one record from fast CCD camera (IDT X-Stream XS3, 4 GB) consisted
first of the 
SSW sweeping so fast that the particle stayed uninfluenced and secondly from the particle motion in stationary SSW.
The first part served for determination of the functions for pattern reconstruction, the second part was used for the particle behavior analyses. 
We used the same frame rate equal to 6120 fps and integration time in the range from 2 to 8 $\mu$s according to the size of the particle. 
CCD camera provided intensity resolution in 1024 levels and the S/N ratio was always better than 100 and therefore we 
do not expect serious errors coming from weak signal detection \cite{CheezumBJ01}.

\begin{figure}[hp]
\scalebox{0.52}{\includegraphics{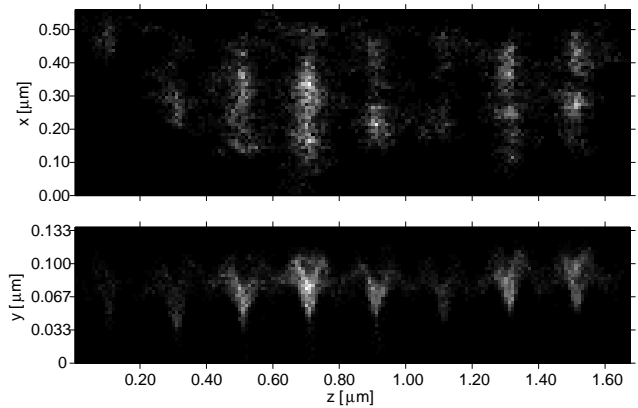}}
%\scalebox{0.4}{\includegraphics{Fig5b.eps}}
\caption{
Histograms of particle (diameter 600~nm) positions. 
The longitudinal trap positions (z) are clearly separated and they are much narrower comparing to the lateral one (x). 
In the direction perpendicular to the prism surface (y) the object closer to the surface is 
longitudinally confined much tightly. 
Despite the very rough position measurements in this axis, this conclusion is quite coherent with the expected exponential decay of the trap depth due to the dominant evanescent field.}
\label{hist}
\end{figure}

In the subsequent experiments we tuned the beam incident angle
so that the transmissive field was negligible and therefore we could
assume exponential decay of the SSW
above the surface. 
Therefore the relative position of the object in this direction we estimated from the ratio of total image intensities of measured and of the best fitted calculated image.
To roughly calibrate this axis we assumed that the velocity distributions in all three dimensions must be the same since the time between 
subsequent measured positions is much smaller to see a significant effect of optical potential landscape on the particle motion. 
An example of the results from these procedures is shown in Fig. \ref{hist} in the form of histograms of $10^4$ particle positions taken within 1.6 sec. 

%Fig. 5

\begin{figure}[ht]                                                               
\scalebox{0.68}{\includegraphics{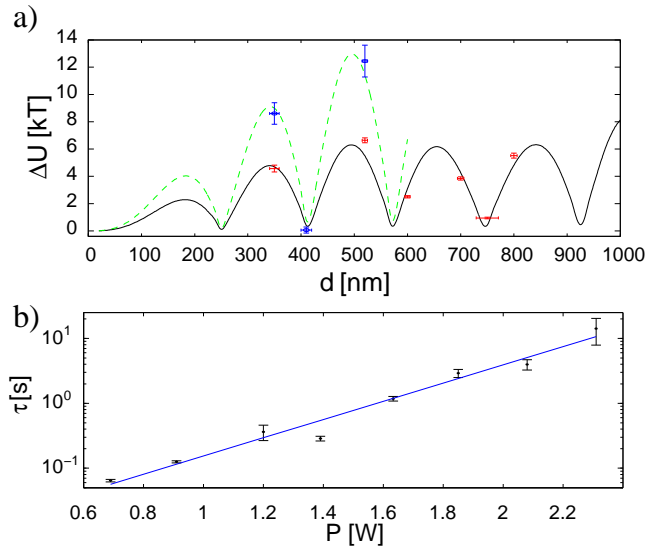}}
\caption{a)
Calculated (full and dashed lines) and measured 
(+ and o) 
depths of the optical traps in longitudinal direction. 
The dashed line is for setup modified with slightly smaller incident angle. The vertical error bars 
coincided with the 95 $\%$ confidential level and the horizontal ones express standard deviation of the size of experimentally used particles. 
b) Dependence of the mean passage time $\tau$ for a polystyrene bead of diameter 520 nm from one trap to neighboring trap in SSW on the power of the trapping laser. 
These values were taken from 100 to 500 jumps and from the fit of exponential function 
$\exp(-t/\tau)/\tau$ to the time $t$ dependence of the distribution of the measured passage times for each power. 
The error bars show the confidential level equal to $95\%$ for the fitted time constants $\tau$.
}
\label{trapdepth}
\end{figure}

We compared 
the trap depth in the longitudinal direction for particles of different sizes. We assumed  
a Boltzmann distribution for each measured particle size \cite{FlorinAPA98} and we fitted a function $a \exp[-b \sin^2(z/\lambda+c)/(k_B T)]$ to the measured histograms with unknown parameters $a,\, b,\,c$ and fixed values of wavelength $\lambda$, temperature $T$, and  Boltzmann constant $k_B$.
We carefully proceeded all the experiments with different particle sizes using exactly the same setup within 2 hour period. 
Even though we cannot measure precisely the optical intensity in the SSW, we are convinced that the optical setup including the optical intensity was the same for all measured particle sizes throughout. 
This enabled us to compare the experimental results with the theoretical model in the following way. 
We calculated the trap depths for the particle sizes used in experiments for different incident angles of the plane waves. 
The plane waves had the same intensity and we looked for the same single multiple of this intensity for all sizes that gave the best coincidence with the experimental results. 
The results for two different arrangements are shown in Fig. \ref{trapdepth}a. 
%Fig6
In spite of the simplicity of the theoretical model the coincidence with the measurement is very good, especially with the sizes of objects unaffected by SSW. 
Since the particle behavior is controlled by stochastic processes of Brownian motion, we measured how the mean passage time from one trap to the neighboring trap depends on the trapping laser power (trap depth). 
Assuming exponential distribution for passage time, the mean passage time also corresponds to the period during which only $100 \exp(-1)\%\simeq 36.8 \%$  of the particles stayed in the same trap. 
Inverse of mean passage time is for one dimensional and bistable case called Kramers rate \cite{KramersPHYS40}. 
The results obtained for the particle of size  $520$ nm are shown in Fig. \ref{trapdepth}b. 

In conclusion, we have demonstrated how two counter-propagating surface fields (combination of  
evanescent  and 
propagating fields) 
can be used to create surface standing wave (SSW) where the sub-micron  
particles may be sorted according to their size in the absence of 
any imposed flow. This system 
 may have potential for the organizing of  
biological and colloidal sub-micrometer objects localized on the surface. 
We quantitatively characterize the depth of the optical traps along the beams and proved that for certain 
object sizes it is much deeper and confine particles for longer time than for the others. 
In dense samples where particles sensitive and insensitive to the SSW were mixed, we observed an optical interaction between both types of 
particles once they were in close proximity to one another.
This in itself has been employed to deliver even insensitive particles with the help of bigger or smaller particles confined in the sliding SSW. 
This form of "optical binding" could find very interesting applications in the field of self-organizing sub-micron objects and 
even living cells.  
The affinity of the particle size to the moving SSW was used to shows how diluted solutions of sub-micrometer colloids can be optically sorted without 
fluid flow by exposing the particles to a suitable optical washboard potential.
Analogous calculations show that sorting may also 
occur due to refractive index in this situation.

This work was partially supported by the EC 6FP NEST ADVENTURE Activity 
(ATOM3D, project No. 508952), ISI IRP (AV0Z20650511), GA ASCR (IAA1065203), ESF EUROCORES programme (through project NOMSAN) with funds from the UK EPSRC.

%\clearpage 
%\bibliography{tweeznew}   %
%\bibliographystyle{apsrev}   %

\end{document}